# Расщепление фотона – двадцать лет спустя

З.К. Силагадзе

Институт Ядерной Физики им. Г.И. Будкера и Новосибирский Государственный Университет

**В 1995 году команда физиков из Института ядерной физики в Новосибирске впервые смогла наблюдать расщепление фотона в поле атомного ядра и доложила предварительные результаты этого эксперимента на двух конференциях. Это чрезвычайно трудный эксперимент, так как вероятность процесса очень маленькая. Потребовалось еще семь лет, чтобы опубликовать окончательные результаты [1]. Недавно эта история получила дальнейшее развитие. Детектор ATLAS на большом адроном коллайдере в ультра-периферийных столкновениях тяжелых ионов зарегистрировал родственный к расщеплению фотона процесс - рассеяние света на свете [2]. Кроме того, команда итальянских, польских и британских астрофизиков получила первое наблюдательное указание на существование вакуумного двулучепреломнения [3] в магнитном поле изолированной нейтронной звезды – физическое явление, тоже родственное к расщеплению фотона.**

Уравнение Максвелла, которые описывают многообразие электромагнитных явлений в классической физике, линейны. Это означает, что кванты электромагнитного поля, фотоны, не взаимодействуют с друг другом – классически свет на свете не рассеивается. Но в квантовой теории такой процесс становится возможным: фотон на короткое время может превратится в виртуальную пару заряженных частиц, и встречный фотон может провзаимодействовать с этой виртуальной парой – в результате получится упругое рассеяние фотона на фотоне (см. правый рисунок).

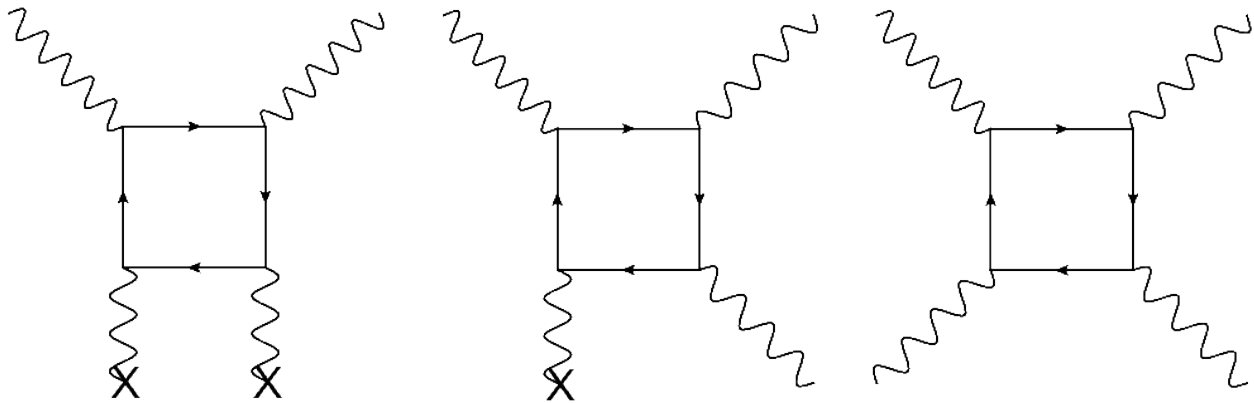

Рассеяние света на свете – фундаментальный процесс в квантовой электродинамике. При этом, если один из фотонов заменяется на внешнее электромагнитное поле (обозначенное крестиком на рисунке), получим родственные процессы: рассеяние Дельбрюка (левый рисунок) и расщепление фотона в электромагнитном поле (центральный рисунок).

Из этих нелинейных процессов квантовой электродинамики, проще всего изучать дельбрюковское рассеяние в кулоновском поле тяжелого ядра, так как его сечение пропорционально четвертой степени заряда ядра и поэтому сравнительно большое. Кроме того сечение дельбрюковского рассеяния сосредоточено в основном в области малых углов рассеяния, что облегчает выделение этого процесса из фона. Например, уже при энергии фотона 300 МэВ и для углов рассеяния порядка 0.01 градуса, сечение дельбрюковского рассеяния превосходит соответствующее сечение комптоновского рассеяния на три порядка.

История дельбрюковского рассеяния [4] начинается в 1933 году когда Лиза Мейтнер и ее аспирант Костер опубликовали экспериментальную работу, в которой изучалось рассеяние гамма-квантов с энергией 2.615 МэВ в свинце и железе. Сечение неупругого рассеяния хорошо соответствовало теоретическому результату Клейна и Нишины для комптоновского рассеяния. Но Мейтнер и Костер наблюдали также значительное упругое рассеяние неизвестной природы. В приложении к их статье, Дельбрюк
предложил объяснение этого явления с помощью моря Дирака. Согласно гипотезе Дирака, которая появилась годом раньше, все уровни с отрицательной энергией, существование которых следует из релятивитского уравнения Дирака, заполнены электронами. Обычно гамма-квант не может взаимодействовать с этим морем дирака, так как результатом взаимодействия был бы процесс, нарушающий законы сохранения энергии и импульса: гамма-квант перебрасывает один из электронов из моря Дирака в область состоянии с положительной энергией и сам исчезает. В результате вместо гамма-кванта получаем пару электрон-позитрон, где позитрон соответствует дырке в море Дирака, которая образовалась, когда из него вырвали электрон. Но согласно Дельбрюку, присутствие ядра искажает море Дирака (по современной терминологии, происходит поляризация вакуума вокруг ядра), и электрон уже может рассеяться на электроне с отрицательной энергией в этом искаженном море Дирака, подобно рэлеевского рассеяния низкоэнергетических фотонов на атоме. При этом энергия гамма-кванта не меняется, так как электрон, на котором он рассеивается, оставаясь в море отрицательных энергий, не может изменить свою энергию, так как *в*се отрицательные уровни энергии заняты, а принцип Паули запрещает нахождение двух электронов в одном и том же квантовом состоянии. Сохранение импульса обеспечивается присутствием ядра, которое испытывает отдачу.

Как вспоминает сам Дельбрюк [5], его описание процесса рассеяния было не совсем верным: из диаграммы Фейнмана, приведенного выше, видно, что гамма-квант не рассеивается на электроне, который все время сидит в море Дирака, а для начала порождает электрон-позитронную пару при взаимодействии с виртуальным фотоном из кулоновского поля ядра. Но тем не менее Дельбрюк взялся вычислять сечение этого процесса, и это вычисление превратилось в кошмар для него. Наконец по совету Бете он предсказал, что сечение этого процесса должно быть пропорционально четвертой степени заряда ядра и на этом его попытки вычисления сечения этого процесса и закончились. Он опубликовал вышеупомянутое как прилоожение к статье Мейтнера и Костера и ничего не слышал об этой проблеме последующие двадцать лет, так как ушел в биологию. В пятидесятые годы кто-то сказал ему, что в журнале Physical Review были опубликованы две статьи Бете и его аспиранта о рассеянии Дельбрюка. Так этот термин (рассеяние Дельбрюка) и вошел в науку с легкой руки Бете. На фотографиях ниже изображены Макс Дельбрюк (слева) и Ганс Бете (справа).

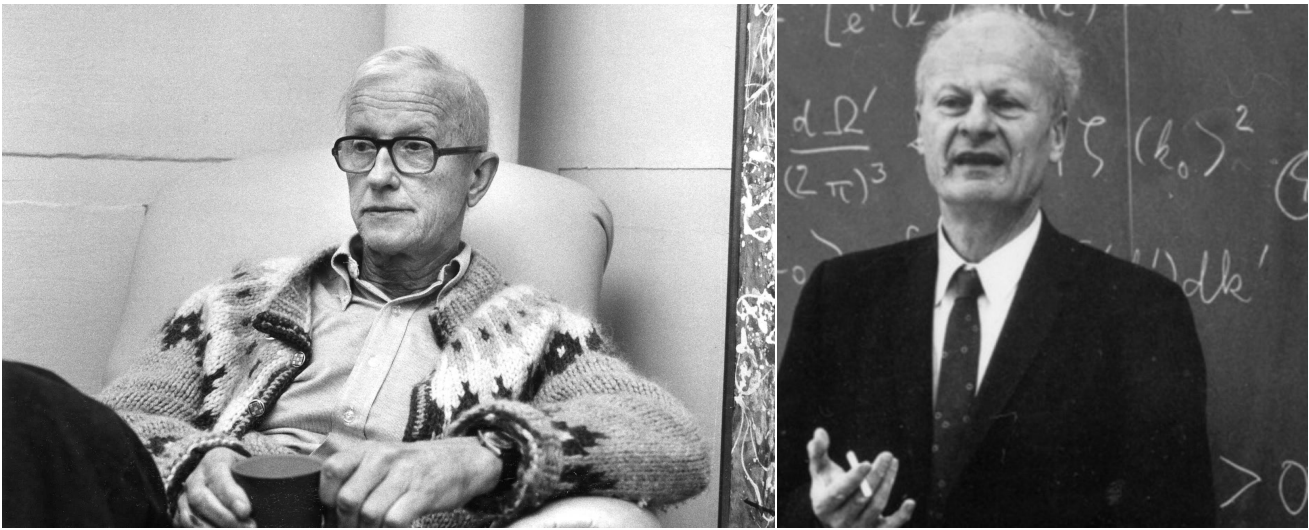

Через несколько месяцев после опубликования заметки Дельбрюка, Халперн предположил, что должны существовать еще два родственных процесса: расщепление фотона в поле ядра и рассеяние фотона на фотоне. Первую попытку вычисления этих процессов предпринял Кеммер в 1937 году, но настоящий прогресс наступил только после развития квантовой электродинамики в ее современном виде. Общий анализ четырехугольной петлевой диаграммы с виртуальным электроном, которая играет важную роль в этих нелинейных процессах квантовой электродимамики, был выполнен в работе Карплуса и Неймана в 1950 году. Через два года Рорлих и Глукштерн впервые вычислили амплитуду дельбрюковского рассеяния вперед. Вычисление было облегчено тем, что мнимая часть амплитуды рассеяние вперед получается из известной амплитуды рождения электрон-позитроной пары с помощью оптической теоремы квантовой теории поля. Реальная часть амлитуды тогда вычисляется из мнимой части с помощью так называемых дисперсионных соотношении. Рорлих и Глукштерн отметили в своей работе, что полное вычисление амплитуды дельбрюковского рассеяния для любых углов в рамках квантовой электродинамики представляет собой чрезвычайно трудную задачу. Поэтому неудивительно, что в последующие двадцать лет прогресс в изучении дельбрюковского рассеяния был очень маленьким. Это относится не только к теории, но и эксперименту.

Перелом наступил в семидесятые и восьмидесятые годы прошлого века, когда были предложены несколько приближенных методов расчета, справедливых в следующих областях параметров: большие энергии и малые углы, малые энергии без ограничения углов, промежуточные энергии и сравнительно большие углы рассеяния. Важную роль в этом прогрессе сыграли Новосибирские физики из института ядерной физики (В.Н. Байер, В.М. Страховенко, В.М. Катков, А.И. Мильштейн), которые под руководством Владимира Байера развили квазиклассический операторный подход к описанию процессов квантовой электродинамики во внешних электромагнитных полях. Преимуществом операторного метода является то, что позволяет вычислить амплитуды процессов без использования явного вида пропагатора электрона во внешнем поле. На фотографии ниже Владимир Николаевич Байер.

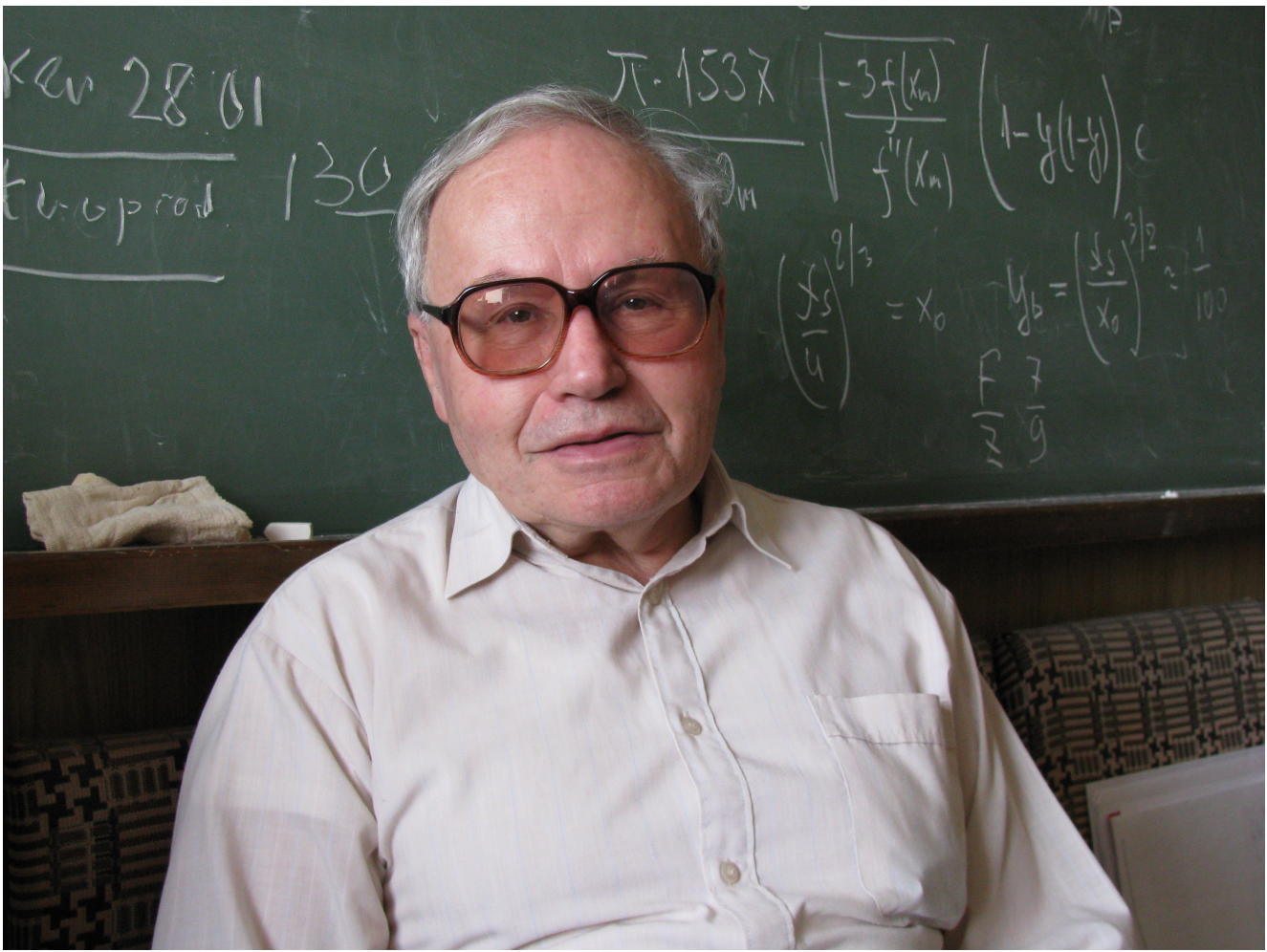

В экспериментальном изучении нелинейных процессов квантовой электродинамики Новосибирский институт ядерной физики тоже сыграл ведущую роль. Несколько обстоятельств способствовали тому, что в этом институте сложились уникальные условия для изучении нелинейных процессов квантовой электродинамики. Во первых в апреле 1992 года заработал электрон-позитронный коллайдер ВЭПП-4М, который на сегодняшний день является самым большим действующим коллайдером в России. Через несколько месяцев вступил в строй установка РОКК-1М, которая позволяет получать интенсивный пучок высокоэнергетичных поляризованных гамма-квантов с энергиями до 1 ГэВ путем рассеяния лазерного излучения на встречном электронном пучке коллайдера ВЭПП-4М. Пионером этой экспериментальной техники в ИЯФ-е был Гурам Кезерашвили, показаный на фотографии ниже.

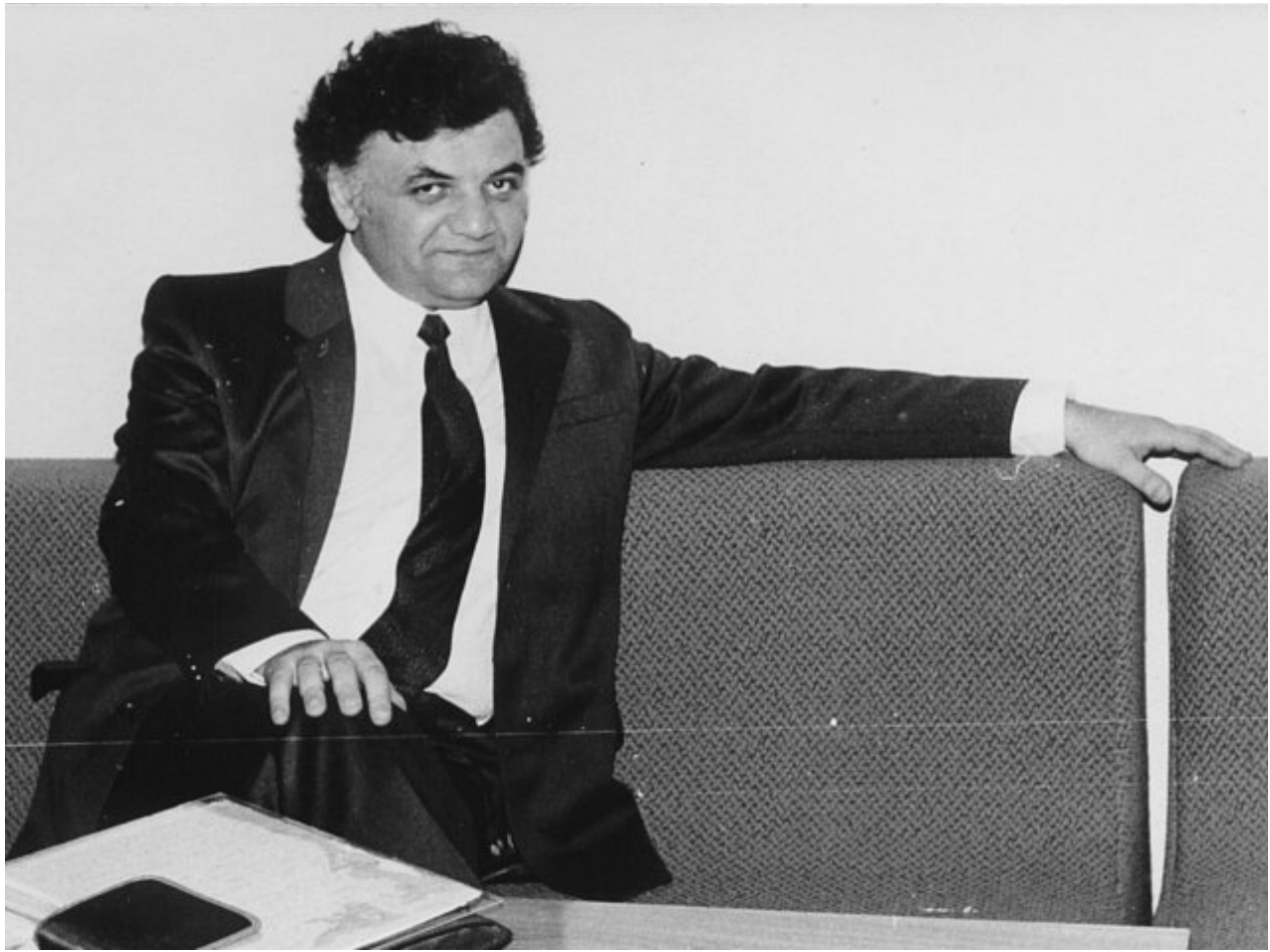

Кроме того на ВЭПП-4М для основных экспериментов на детекторе КЕДР была создана система регистрации рассеянных электронов, который позволяет измерить энергию каждого рассеянного фотона: при взаимодействии лазерного излучения с пучком электронов, рассеянный фотон вылетает из вакуумной камеры коллайдера по направлению импульса электрона, а сам электрон отклоняется поворотным магнитом и попадает в систему регистрации рассеянных электронов, где определяется отданная фотону энергия.

В 1994-1997 годах на установке РОКК-1М был проведен эксперимент по изучению двух нелинейных процессов квантовой электродинамики: дельбрюковского рассеяния и расщепления фотона в кулоновском поле ядра. На фотографии ниже показана часть экспериментальной установки [6].

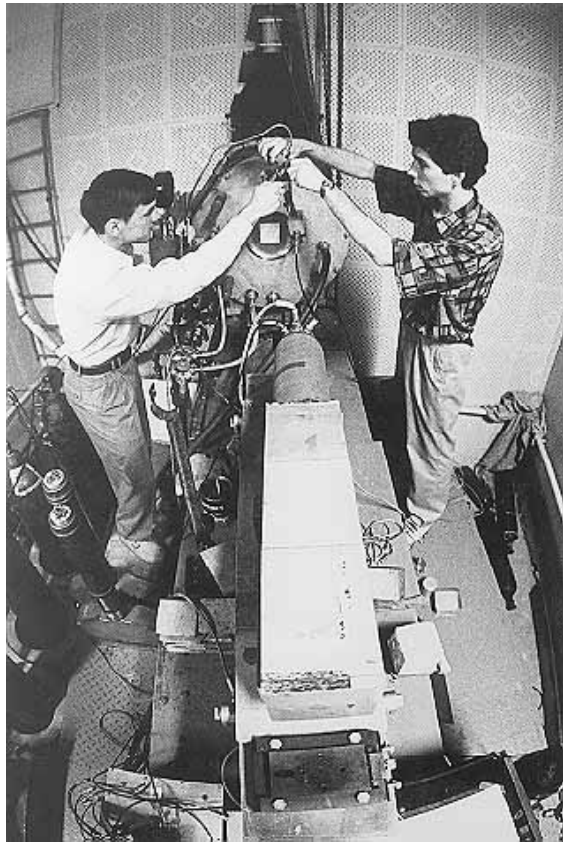

Рассеянный фотон (гамма-квант) с энергией от 100 до 450 МэВ попадал на мишень из кристалла германата висмута, расположенную на расстоянии около 30 метров от точки рассеяния лазерного фотона. Кристалл германата висмута, специально изготовленный для этого эксперимента в институте неорганической химии СО РАН, играл двоякую роль. С одной стороны он был мишенью содержащую тяжелые элементы с большим зарядом ядра, а с другой стороны он сам был детектором: регистрировал выделение энергии ионизации в процессе взаимодействия гамма-кванта с мишенью, если эта энергия выделялась (в искомых процессах энергия как раз не должна была выделяться). На расстоянии около 5 метров за мишенью был расположен ионизационный калориметр на основе 400 кг жидкого криптона – детектор рассеянных фотонов.

События дельбрюковского рассеяния и расщепления фотона в этом эксперименте выглядели так: система регистрации рассеянных электронов регистрировал один электрон и по потере энергии этого рассеянного на лазерном фотоне электрона измерялся энергия первоначального гамма-кванта. В случае дельбрюковского рассеяния, ионизационный калориметр (детектор фотонов) регистрировал один фотон с такой же энергией, как первоначальный гамма-квант, а в случае расщепления фотона регистрировались два фотона, сумма энергии которых равнялась энергии первоначального гамма-кванта. В обеих случаях, отсутствовало энерговыделение в кристалле германата висмута.

Из трех миллиардов первоначальных гамма-квантов около десяти тысяч испытали дельбрюковское рассеяние и около 150 расщепились в поле ядра. В результате эксперимента, более чем на порядок была улучшена экспериментальная точность измерения основных параметров дельбрюковского рассеяния в области высоких энергий. Для объяснение результатов эксперимента потребовались более точные расчеты с учетом так называемых кулоновских поправок. Расщепление фотона экспериментально наблюдалось впервые.

Интересно, что существовала надежда, что рассеяние Дельбрюка, помимо чисто академического интереса, может иметь и практические применения. При измерении показателя преломления кремния для гамма-квантов МэВ-них энергий вдруг оказалось [7], что этот показатель преломления больше единицы, тогда как из теории следовало, что этот показатель, который должен был определяться релеевским рассеянием фотонов в кристалле, меньше единицы. Авторы работы объясняют свой результат вкладом дельбрюковского рассеяния и их оценки с учетом такого вклада хорошо согласуются с экспериментальными результатами. То, что вклад дельбрюковского рассеяния может пересилить вклад релеевского рассеяния в реальной части показателя преломнения и привести изменению знака полного вклада, было удивительно и встречено с некоторым недоверием (см. Здесь [8] и здесь [9]). Если бы этот экспериментальный результат подтвердился, обнаруженный эффект открыл бы новую область исследований, а именно гамма-оптику со многими потенциальными применениями. К сожалению, оказалось [10], что систематические ошибки исказили интерпретацию экспериментального результата, а новые эксперименты не подтвердили наличие эффекта [11].

Для оптических фотонов сечение рассеяния света на свете настолько мало, что казалось бы нет никакого шанса зарегистрировать его в лаборатории (см. однако вот эту статью [12]). Недавно был придуман [13] весьма хитроумный способ наблюдения рассеяния света на свете в ультра-периферийных столкновениях тяжелых ионов. Такие столкновения происходят с большим прицельным параметром и поэтому сами ядра непосредственно с друг другом не взаимодействуют. Но их электромагнитные поля, которые приближенно можно представить как поток почти реальных (квазиреальных) фотонов, могут взаимодействовать с друг другом и в частности могут привести к рассеянию света на свете по механизму показанному ниже.

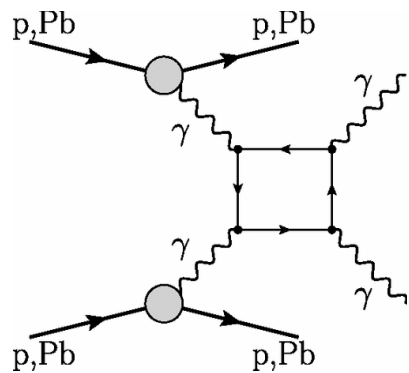

Такие события исключительно легко распознать в детекторе, так как они почти не имеют фона. Обычно, при жестком столкновении тяжелых ионов, детектор регистрирует около тысячи вторичных частиц, тогда как при рассеянии света на свете регистрируются только два фотона с противоположными поперечными импульсами и больше ничего.

Поток эквивалентных фотонов, сопровождающий ультра-релятивисткое ядро, пропорционален квадрату заряда ядра. Поэтому в столкновениях ядер свинца с друг другом общий поток пропорционален четвертой степени заряда ядра свинца, и это очень большое число – около пятьдесят миллионов. Именно благодаря этому усилению эффекта и удалось наблюдать рассеяние света на свете [2], который сам по себе очень редкий процесс. В почти четырех миллиардах столкновениях ядер свинца с друг другом, только 13 событий возможного рассеяния света на свете были зарегистрированы. В пределах погрешности эксперимента это соответствует тем ожиданиям, которые следуют из квантовой электродинамики и теории сильного взаимодействия для основного (ожидалось 7.3 сигнальных событий) и фоновых (ожидалось 2.6 фоновых событий) процессов. Основный источник фона – это когда ядра вместо фотонов излучают глюоны, переносчики сильного взаимодействия, в бесцветном состоянии и они в свою очередь, через кварковую петлю, порождают два фотона. Но ядра, с точки зрения сильных взаимодействии, очень рыхлые объекты – их энергия связи всего лишь порядка 8 МэВ. Поэтому, как правило, излучение даже мягкого глюона сопровождается выбиванием из ядра нескольких нуклонов и эти нуклоны могут зарегистрироваться в так называемых передних детекторах ATLAS-а. Это позволяет существенно подавить данный фон.

Если в диаграмме, который описывает рассеяние Дельбрюка, заменить внешнее кулоновское поле на внешнее магнитное поле, окажется, что разные состояние поляризации света распространяются в магнитном поле с разной скоростью – этот эффект носит название вакуумного двулучепреломления. Для достижимых в лаборатории интенсивностей магнитных полей, это очень слабый эффект, и хотя проводятся эксперименты с целью обнаружения вакуумного двулучепреломления в лаборатории, например эксперимент PVLAS [14] (см. фотографию ниже), до сих пор обнаружить этот эффект не удавалось. Но помогли астрофизические наблюдения.

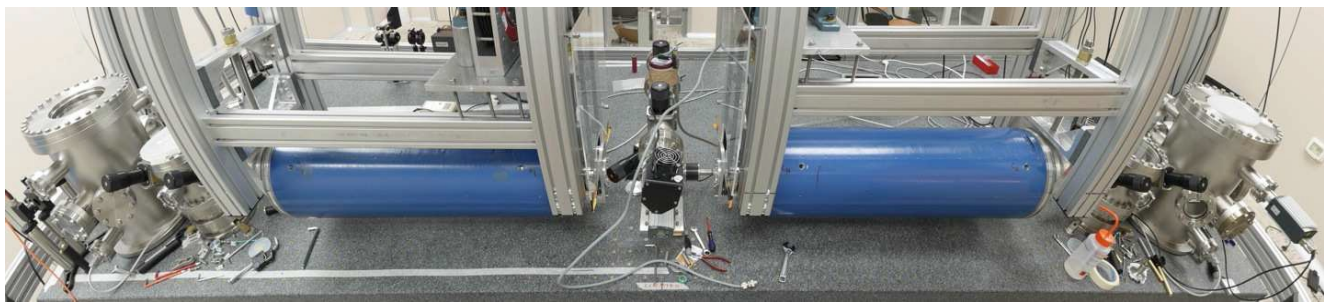

В девяностых годах прошлого века обнаружили «великолепную семерку» изолированных нейтронных звезд, которые не излучали в радио диапазоне, а их излучение носило полностью тепловой характер и исходило от поверхности звезд. Из наблюдаемых свойств этого излучения следовало, что звезды обладают очень сильным магнитным полем, порядка $10^{13}$ Гаусс. В таких сильных полях тепловое излучение должно быть поляризованным, но из-за того, что на Земле мы получаем излучение со всей поверхности звезды, а вдоль поверхности направление магнитного поля сильно меняется, наблюдаемая на Земле поляризация должна была усредняться практически до нуля. Ситуация драматически меняется если учтем вакуумное двулучепреломления, которое в таких сильных полях уже существенный эффект. В зависимости от геометрии наблюдения и от механизма поверхностного излучения, степень наблюдаемой линейной поляризации за счет эффекта вакуумного двулучепреломления может увеличится почти до ста процентов.

Недавно ученые из Италии, Польши и Великобритании измерили степень линейной поляризации оптического излучения звезды RX J1856.5-3754 [3] из «великолепной семерки». Основная часть излучения звезды приходится на мягкий рентген, и в этой области спектра пока поляризационные измерения не доступны. Но с помощью Очень Большого Телескопа (VLT – Very Large Telescope – он состоит из четырех громадных оптических телескопов диаметром 8.2 м каждый, см. фотографии ниже), расположенного в обсерватории Параналь в Чили, ученым удалось измерить поляризацию слабой оптической компоненты излучения. Степень поляризации оказалось большой – около шестнадцати процентов. По мнению авторов работы, такая большая степень наблюдаемой поляризации является первым наблюдательным указанием на существование вакуумного двулучепреломления и тем самим подтверждает предсказания квантовой электродинамики в сильных полях.

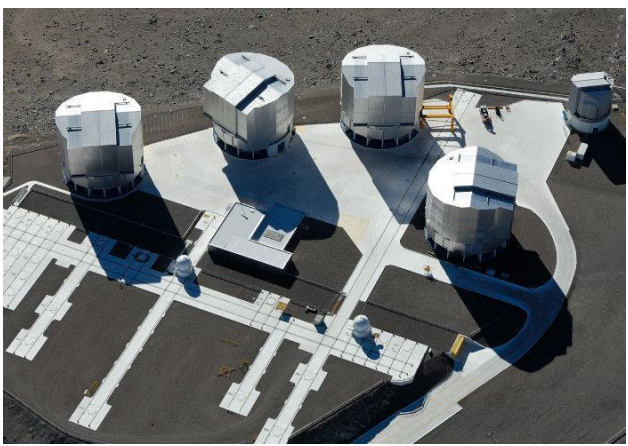
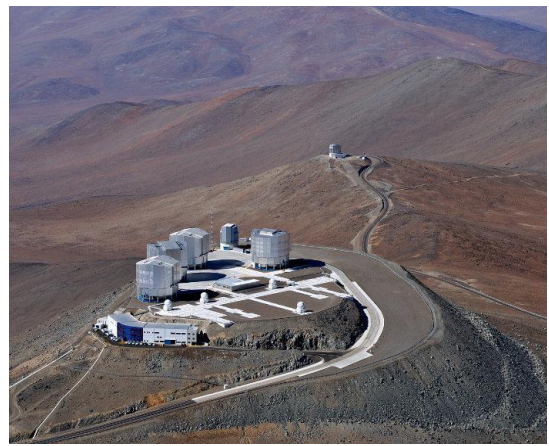

## Список цитированной литературы